\documentclass[a4paper,10pt,aps,pre,twocolumn,superscriptaddress,amsmath,amssymb,nofootinbib]{revtex4}
\usepackage{graphicx,color,soul}

\newcommand{\mean}[1]{\langle #1 \rangle}

\renewcommand{\vec}[1]{\mathbf #1}

\newcommand{\al}{\alpha}

\newcommand{\vhi}{\varphi}
\newcommand{\sig}{\sigma}

\newcommand{\x}{\vec r}
\newcommand{\Dr}{D_\text{r}}

\newcommand{\pb}{P_\text{B}}
\newcommand{\pbm}{P_\text{B}^\text{m}}
\newcommand{\nois}{\boldsymbol\xi}
\newcommand{\deff}{d_\text{eff}}
\newcommand{\tx}{\tau_\text{x}}
\newcommand{\tf}{\tau_\text{f}}

\begin{document}

\title{Nucleation pathway and kinetics of phase-separating active Brownian particles}

\author{David Richard}
\affiliation{Institut f\"ur Physik, Johannes Gutenberg-Universit\"at Mainz, Staudingerweg 7-9, 55128 Mainz, Germany}
\author{Hartmut L\"owen}
\affiliation{Institut f\"ur Theoretische Physik II, Heinrich-Heine-Universit\"at D\"usseldorf, Universit\"atsstra{\ss}e 1, 40225 D\"usseldorf, Germany}
\author{Thomas Speck}
\affiliation{Institut f\"ur Physik, Johannes Gutenberg-Universit\"at Mainz, Staudingerweg 7-9, 55128 Mainz, Germany}

\begin{abstract}
  Suspensions of purely repulsive but self-propelled Brownian particles might undergo phase separation, a phenomenon that strongly resembles the phase separation of passive particles with attractions. Here we employ computer simulations to study the nucleation kinetics and the microscopic pathway active Brownian disks take in two dimensions when quenched from the homogeneous suspension to propulsion speeds beyond the binodal. We find the same qualitative behavior for the nucleation rate as a function of density as for a passive suspension undergoing liquid-vapor separation, suggesting that the scenario of an effective free energy also extends to the kinetics of phase separation. We study the transition in more detail through a committor analysis and find that transition states are best described by a combination of cluster size and the radial polarization of particles in the cluster.
\end{abstract}

\maketitle


\section{Introduction}


Understanding the transformation of matter in response to a change of external conditions such as temperature and pressure is one of the cornerstones of statistical physics. Recently, the notion ``matter'' has come to include ``active matter''~\cite{rama10,marc13} covering a broad class of systems in which autonomous constituents convert (free) energy into directed motion, ranging from flocks of birds~\cite{cava10} to bacteria~\cite{wens12}. Of particular interest are suspensions of self-propelled colloidal particles~\cite{bial14} due to their potential applications, \emph{e.g.}, for self-assembly~\cite{kumm15,grun15,meer16} and sorting~\cite{mija13}.

Active Brownian particles are a minimal model for active matter combining volume exclusion with directed motion but neglecting long-ranged phoretic and hydrodynamic interactions. For sufficiently high densities and driving speeds, active Brownian particles undergo ``clustering''~\cite{yaou12,redn13,butt13,sten13,sten14,wyso14,zott14}, which strongly resembles the macroscopic liquid-vapor phase separation of passive Brownian particles. This phenomenon has been described as a ``motility-induced phase transition''~\cite{cate13,cate15} caused by the self-blocking of particles due to the interplay between directed motion and volume exclusion, the microscopic picture of which has been confirmed in experiments~\cite{butt13}. Starting from the microscopic many-body dynamics, one can indeed systematically derive an effective free energy for the density~\cite{bial13,spec14,spec15}, which links active Brownian particles to the scenario of liquid-vapor separation. However, such an effective description is valid only on length scales larger than the persistence length of the directed motion~\cite{spec15}.

For short-ranged interactions, liquid-vapor phase separation of passive Brownian
particles falls into the universality class of the Ising model and has been studied extensively~\cite{bind87}. Below the critical temperature it is a first-order transition accompanied by hysteresis, which can be understood from the competition between the gain of bulk free energy driving the transition and the penalty that is associated with the formation of an interface between stable and metastable phase. As a result, the nucleation rate shows a typical non-monotonic behavior. For fixed temperature it vanishes at the ``binodal'' (given by the coexisting densities) and increases with increasing density (supersaturation), indicating a decreasing free energy barrier. The region where this barrier becomes of the order of the thermal energy is typically identified with the ``spinodal'' (which becomes an exact line only in mean-field models). Increasing the supersaturation further leads to a reduced collective diffusion and consequently the nucleation rate also decreases.

Active Brownian disks also show nucleation behavior in the vicinity of the binodal~\cite{redn13,spec14}. Here we demonstrate numerically that the nucleation rate shows the same qualitative behavior as expected for liquid-vapor separation. Our analysis reveals that the transition is characterized by a well-defined transition state separating the metastable homogeneous suspension (``reactants'') from the stable phase-separated state (``products''). For systems governed by dynamics that obey detailed balance, such a behavior is typically rationalized in terms of a free energy barrier that has to be overcome by a rare, highly collective fluctuation. In agreement with this picture, we find that when decreasing the density, the nucleation rate becomes so small that it is not observed anymore in direct simulation runs. To circumvent the prohibitively long time scales, one has to resort to numerical rare-event methods, which is an active field in computational chemistry~\cite{dell02,e02,erp05}. Here we employ Forward Flux Sampling~\cite{alle06,alle09}, which can also be used for dynamics that explicitly break detailed balance~\cite{alle08} as is the case for active Brownian particles. This allows us to obtain rates that span more than ten orders of magnitude.

Finally, to gain further insights into the transition state~\cite{humm04} and the nucleation pathway, we perform a committor analysis~\cite{dell02,prin11a} in the intermediate regime where nucleation is rare but still accessible by direct simulations. Committor analyses have become a standard tool in computational chemistry used, \emph{e.g.}, to describe the demixing of a binary mixture~\cite{scho10} and crystallization~\cite{jung11,lech11}, for example of water~\cite{haji15} and hard spheres under shear~\cite{rich15}. We describe three order parameters that capture different microscopic aspects and construct a reaction coordinate through likelihood maximization~\cite{pete06,lech11,jung13}, which indicates that the polarization of particles plays a role in the nucleation of dense domains of active Brownian disks.


\section{Model}

We study suspensions of $N$ disks in two dimensions with periodic boundary conditions. The position of disk $k$ is $\x_k$ with orientation $\vec e_k=(\cos\vhi_k,\sin\vhi_k)^T$ determined by the angle $\vhi_k$. A single configuration $x=\{\x_k,\vhi_k\}$ is thus given by the positions and orientations of all disks. Particles interact pairwise through the short-ranged, repulsive Weeks-Chandler-Andersen potential
\begin{equation}
  \label{eq:wca}
  u(r) = 4\epsilon\left[ (\sig/r)^{12} - (\sig/r)^6 + \frac{1}{4} \right]
\end{equation}
for $r<2^{1/6}\sig$ and zero otherwise, with (dimensionless) potential strength $\epsilon=100$. Disk positions evolve according to
\begin{equation}
  \label{eq:lang}
  \dot\x_k = -\nabla_kU + v_0\vec e_k + \nois_k,
\end{equation}
where the noise $\nois_k$ models the interactions with a heat reservoir assuming correlations $\mean{\xi_{ki}(t)\xi_{lj}(t')}=2\delta_{kl}\delta_{ij}\delta(t-t')$ and zero mean. The conservative force stems from the potential energy $U(\{\x_k\})=\sum_{k<l}u(|\x_k-\x_l|)$. The direction motion is modelled by an effective force $v_0\vec e_k$ along the orientation of particles with free swimming speed $v_0$. In addition, the orientations undergo free rotational diffusion with rotational diffusion coefficient $\Dr=3D_0/\deff^2$ modeling the no-slip boundary condition as is appropriate for colloidal particles. Here, $\deff\simeq1.10688\sig$ is the effective (passive) disk diameter computed via the Barker-Henderson approximation~\cite{bark67}. Moreover, it defines the packing fraction $\phi=\bar\rho\pi(\deff/2)^2$, where $\bar\rho=N/A$ is the global number density with box area $A$. Throughout, we employ dimensionless quantities and measure lengths in units of $\sig$ and time in units of $\sig^2/D_0$, where $D_0$ is the bare translation diffusion coefficient.

\begin{figure*}[t]
  \centering
  \includegraphics{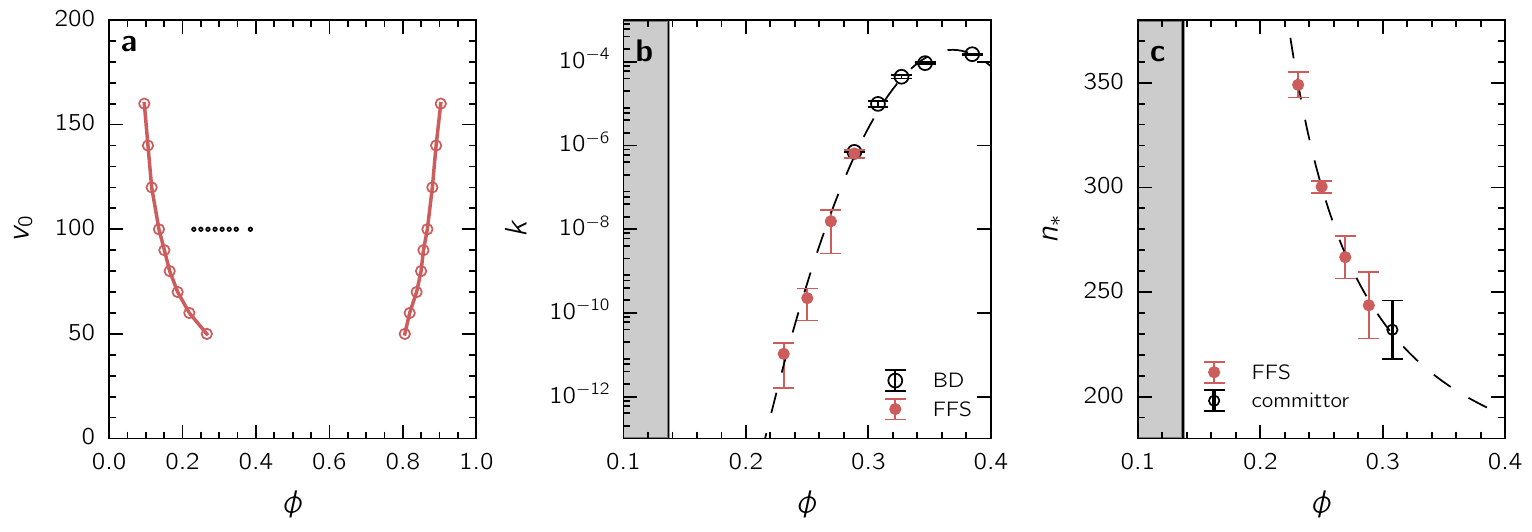}
  \caption{Nucleation kinetics for $N=4096$. (a)~Phase diagram showing the coexisting packing fractions $\phi_\pm$ as a function of swimming speed $v_0$. Small symbols correspond to the state points studied for the nucleation kinetics. (b)~Nucleation rate $k$ \emph{vs.} effective packing faction $\phi$ for $v_0=100$. Black and red circles are computed by Brownian Dynamics (BD) and Forward Flux Sampling (FFS), respectively. The dashed line is a guide to the eye (quadratic fit). (c)~Critical nucleus sizes $n_\ast$. Red symbols are extracted from transfer probabilities in FFS, the black open symbol is the result from an extensive committor analysis of BD trajectories. The dashed line is Eq.~(\ref{eq:ncrit}). The gray areas in (b,c) correspond to the stable homogeneous gas phase for packing fractions $\phi<\phi_-\simeq0.14$. Error bars in (b,c) correspond to the standard error of three independent FFS calculations.}
  \label{fig:kinetics}
\end{figure*}

The equations of motion are integrated with time step $2\times10^{-6}$. Figure~\ref{fig:kinetics}(a) shows the coexisting densities $\phi_\pm$ defining the binodal bounding the two-phase region within which phase separation is possible. The coexisting densities have been obtained from sampled density profiles at area fraction $\phi\simeq0.55$ in an elongated box, see Ref.~\citenum{bial15} for details.


\section{Kinetics}

\subsection{Nucleation rate}

For the kinetics we study $N=4096$ particles varying the packing fraction $\phi$ at fixed swimming speed $v_0=100$. Indicated in Fig.~\ref{fig:kinetics}(a) are the metastable state points for which we have estimated the nucleation rates $k$ shown in Fig.~\ref{fig:kinetics}(b) employing both direct Brownian dynamics (BD) simulations and forward flux sampling (FFS). For both methods initial homogeneous configurations of non-overlapping disks are generated using the algorithm by Clarke and Wiley~\cite{clar87} (at $v_0=0$). The suspension is then quenched instantaneously to $v_0=100$. At sufficiently high supersaturation, the system spontaneously forms many dense domains that evolve towards a single dense domain through coalescence.

\begin{figure}[t]
  \centering
  \includegraphics{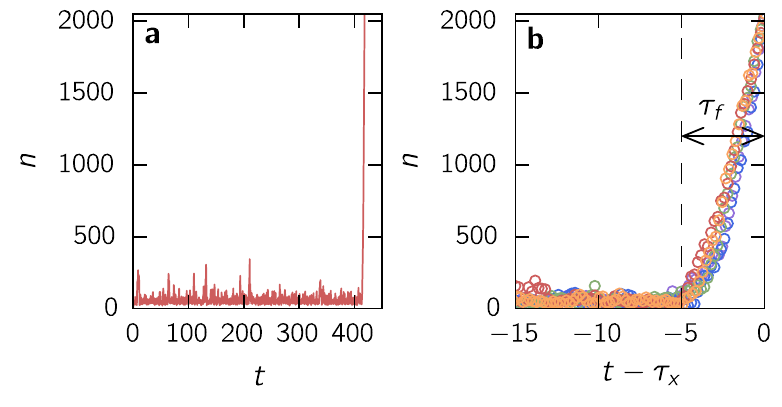}
  \caption{(a)~Single nucleation event for packing fraction $\phi\simeq0.31$. Shown is the size $n$ of the largest cluster. There is a clear separation between induction followed by the growth of the cluster. (b)~Growth behavior of 5 independent runs, where time is shifted by the nucleation time $\tx$ (time to reach $n=2048$ particles). The growth velocities are very similar with $\tf\simeq 5$ to reach the threshold.}
  \label{fig:event}
\end{figure}

Reducing the packing fraction, only a single domain appears after a waiting time (the nucleation time), which then rapidly grows until the system reaches a final steady state. In this regime we employ straightforward BD simulations to obtain the nucleation times $\tx$. To this end we run 50 independent trajectories for each packing fraction and monitor the number and sizes of clusters as a measure for dense domains. Clusters are constructed from mutually bonded particles, where a bond is formed when two disks interact, \emph{i.e.}, if their distance $r$ is less than $r<2^{1/6}\sig$. The nucleation time $\tx$ is stored when the largest cluster with size $n$ reaches the threshold of $n\geqslant2048$, which is much larger than the critical nucleus sizes for the densities studied by direct BD simulations. Since growth is much faster, the actual value of the threshold does not influence significantly the values of $\tx$, cf. Fig.~\ref{fig:event}. The nucleation rate is then estimated from the mean nucleation time through
\begin{equation}
  \label{eq:k:bd}
  k = \frac{1}{\mean{\tx}A},
\end{equation}
where $A$ is the area of the simulation box and $\mean{\cdot}$ denotes the average over sampled trajectories.

\subsection{Forward Flux Sampling}

As the packing fraction $\phi$ is reduced, the nucleation rate $k$ drops. Below $\phi\lesssim0.29$ no formation of dense clusters is observed anymore in direct BD simulations. To estimate the nucleation rate we thus have to resort to FFS. To this end we use the size of the largest cluster and define $M$ interfaces with cluster sizes $n_i$. Configurations that have $n<n_1$ are designated ``basin A'' (the homogenous state) and configurations with largest cluster $n>n_M$ belong to ``basin B'' (phase-separated). The interfaces then guide the system from A to B by running trajectories and sequentially collecting configurations at every interface.

We employ as few interfaces as possible so that conditional probabilities $P(n_{i+1}|n_i)$ for the transitions $i\to i+1$ are approximately independent but there is still a substantial number of trajectories reaching the next interface. The position of the last interface $n_M$ is chosen to have a numerical transfer probability of unity in the forward direction, meaning the irreversible growth of the nucleated cluster. The position of the first interface $n_1$ is determined from the distribution of cluster sizes (obtained from direct BD runs) so that the probability $\sim10^{-3}$ to observe clusters with size $n_1$ is small but not too small. Values range from $n_1=100$, $n_M=600$ and $M=6$ interfaces for the highest packing fraction to $n_1=50$, $n_M=800$ and $M=16$ for the lowest packing fraction studied.

The nucleation rate can be estimated from these simulations through
\begin{equation}
  \label{eq:k:ffs}
  k = k(1|\text A)P(\text B|\text A) = k(1|\text A)\prod_{i=1}^{M-1} P(n_{i+1}|n_i).
\end{equation}
Here, $k(1|\text A)$ is the rate to reach the first interface from the homogenous initial condition. This rate is multiplied by the probability to reach B, which, assuming uncorrelated interfaces, can be expressed as the product of conditional probabilities $P(n_{i+1}|n_i)$ that trajectories reach the next interface $i+1$ when started from interface $i$. We use $50$ independent runs to evaluate $k(1|\text A)$ and collect a set of $500$ configurations at $n_1$. In Fig.~\ref{fig:kinetics}(b) it can been seen that FFS interpolates correctly the BD data to lower packing fractions. To estimate the error we actually perform three independent FFS calculations and determine the standard error.

Moreover, from the transfer probabilities of FFS we extract the critical nucleus size $n_\ast$, see Fig.~\ref{fig:kinetics}(c). For this purpose we need the probabilities
\begin{equation}
  P_j = \prod_{i=j}^{M-1} P(n_{i+1}|n_i)
\end{equation}
to reach B from interface $j$. We model the dependence of this function on cluster size by
\begin{equation}
  \label{eq:tanh}
  P(n) = \frac{1}{2}[1+\tanh(cn+d)]
\end{equation}
with two parameters $c$ and $d$ fitted to the scatter data $(n_j,P_j)$. The critical nuclei obey $P(n_\ast)=\frac{1}{2}$, the value of which are shown in Fig.~\ref{fig:kinetics}(c) as a function of packing fraction $\phi$. We observe that, as the packing fraction decreases, the critical nucleus size increases. The critical sizes are well described by the functional form
\begin{equation}
  \label{eq:ncrit}
  n_\ast(\phi) = n_0 + \frac{b}{(\phi-\phi_-)^{3/2}}
\end{equation}
with fitted offset $n_0\simeq148$ and coefficient $b\simeq5.8$. For passive suspensions, classical nucleation theory~\cite{debenedetti} predicts a function $n_\ast\propto|\Delta\mu|^{-2}$, where $\Delta\mu\propto\phi_--\phi$ is the difference of chemical potential between liquid and vapor. Both functions diverge as the packing fraction $\phi$ approaches the binodal at $\phi_-\simeq0.14$. However, the exponent for active Brownian disks is smaller such that the divergence at coexistence is less severe for active systems. Even more intriguing, whereas for a passive suspension the critical cluster size approaches zero, for active Brownian disks it seems to approach the constant offset $n_0$. BD runs for packing fraction $\phi\simeq0.4$ show that the suspension immediately starts to form dense domains, which indicates that it has become linearly unstable. Still, the size of the critical cluster seems to be $n_\ast\sim200$ particles. This is not necessarily a contradiction since the apparent critical size is now a value that is reached by typical fluctuations, which are much stronger in the driven active system. Hence, at least the critical cluster size departs from the classical picture.

\begin{figure*}[t]
  \centering
  \includegraphics{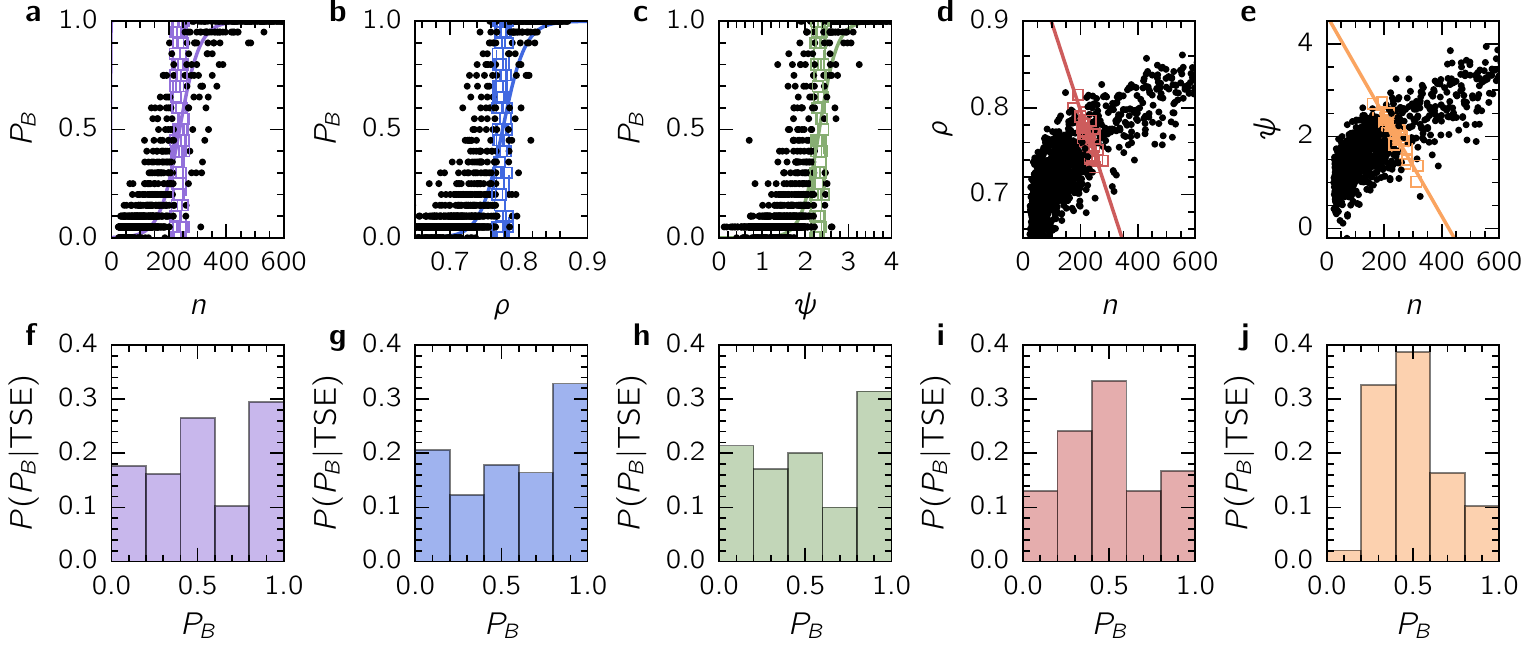}
  \caption{Transition state ensemble (TSE). (a-c)~Committor probability $\pb$ \emph{vs}. the considered order parameters: (a)~size $n$, (b)~density $\rho$, and (c)~polarization $\psi$ of the largest cluster. Solid lines show $\pbm(r)$ [Eq.~(\ref{eq:pbm})]. Colored squares are configurations close to the TSE using the criterion Eq.~(\ref{eq:tse}). (f-h)~Associated distributions $P(\pb|\text{TSE})$ of $\pb$ at the TSE. (d,e)~Results of the log-likelihood maximization for the two models with two variables: (d)~$\{n,\rho\}$ and (e)~$\{n,\psi\}$. (i,j)~Associated distributions of $\pb$. The number of TSE configurations is: 68, 73, 70, 54, 49 (from left to right).}
  \label{fig:tse}
\end{figure*}

\subsection{Committor analysis}

To validate the results for the critical nuclei and to gain further insights into the microscopic pathway how the suspension transforms from homogeneous to phase-separated, we perform a detailed committor analysis for the packing fraction $\phi\simeq0.31$. This is the highest packing fraction that still shows a clear time separation between the appearance of a critical cluster and its growth, cf. Fig.~\ref{fig:event}(a). For any reaction, the committor probability $\pb(x)$ is the exact reaction coordinate. It quantifies the probability to reach basin B from configuration $x$ with isocommittor surfaces $\pb(x)=\text{const}$ in configuration space. In particular, configurations for which $\pb(x)\simeq\tfrac{1}{2}$ constitue the transition state ensemble (TSE).

For the committor analysis we use the 50 trajectories harvested for calculating the rate. From every stored configuration $x_l$ along these trajectories, $N_l$ short ``fleeting'' trajectories are generated. The length $\tf$ of these trajectories is fixed and set to $\tf=5$, which corresponds to the mean time to reach basin B for reactive trajectories, see Fig.~\ref{fig:event}(b). The committor probability is then estimated as the ratio of reactive fleeting trajectories (having reached the threshold) to the total number of fleeting trajectories for that configuration. For the TSE we collect all configurations for which $0.45\leqslant\pb(x)\leqslant0.55$ and calculate the average cluster size $\mean{n}_\text{TSE}=232\pm14$ of these configurations. As shown in Fig.~\ref{fig:kinetics}(c), this value agrees well with the FFS results and continues the trend.


\section{Transition pathway}

\subsection{Reaction coordinate}

The full committor function $\pb(x)$ is costly to determine and does not yield insights into the microscopic details of the transition \emph{per se}. It is often more useful to approximate $\pb$ in terms of a few order parameters, functions $q_i(x)$ that project high-dimensional configurations onto a single number. These order parameters should provide a simplified, low-dimensional description of the progress of the ``reaction'', \emph{i.e.}, the transition from basin A to basin B. In particular, they should allow to construct a good reaction coordinate $r(x)=r(\{q_i(x)\})$. Because the committor probability $\pb(x)$ is the exact reaction coordinate, the isosurfaces of $r(x)$ should closely approximate the isocommittor surfaces, at least close to the TSE corresponding to the value $r_\ast$. For a good reaction coordinate $r$, the distribution $P(\pb|\text{TSE})$ of values $\pb(x)$ for configurations $x$ with $r(x)\simeq r_\ast$ will be strongly peaked around $\tfrac{1}{2}$.

Since the exact committor probability $\pb(x)$ is an unknown function, we need to define a model. A common choice is [cf. Eq.~(\ref{eq:tanh})]
\begin{equation}
  \label{eq:pbm}
  \pbm(r) = \frac{1}{2}(1+\tanh r)
\end{equation}
with linear \emph{ansatz}
\begin{equation}
  \label{eq:r}
  r(x;\{\al_i\}) = \al_0 + \sum_i \al_i q_i(x)
\end{equation}
for the reaction coordinate~\cite{pete06,lech10,lech11}. Without loss of generality, we set $r_\ast=0$. At least close to the TSE such a linear Taylor expansion of a potentially more complicated, non-linear function $r(\{q_i\})$ should be sufficient.

\subsection{Likelihood maximization}

Likelihood methods for the determination of reaction coordinates for nucleation have been developed in Refs.~\citenum{pete06,lech10,lech11} in the context of importance sampling of shooting points. Here we follow the similar approach of Ref.~\citenum{jung13} using as data the fleeting trajectories harvested for the committor analysis. For every stored configuration $x_l$ we run $N_l$ short trajectories of which $n_l$ reach a stable cluster configuration ($x_l\to\text B$) and $N_l-n_l$ fall back to the homogeneous suspension ($x_l\to\text A$). The probability to observe a given sequence of numbers $n_l$ is then given by the likelihood function
\begin{equation}
  \label{eq:L}
  L(\{\al_i\}) = \prod_l {N_l\choose n_l} [\pbm(r_l)]^{n_l}[1-\pbm(r_l)]^{N_l-n_l}
\end{equation}
with values $r_l=r(x_l;\{\al_i\})$ for the reaction coordinate, for which we use Eqs.~(\ref{eq:pbm}) and~(\ref{eq:r}).

Practically, we maximize $\ln L$ using an iterative Newton optimization method to obtain the coefficients $\al_i$ in Eq.~(\ref{eq:r}). Moreover, we employ a bootstrap method to compute the variance of the maximal log-likelihood for each model by resampling our data adding normal noise. The noise variance is chosen to be equal to the variance of the residual between our data and the model extracted from the first maximization. We find that all combinations have values for the log-likelihood that are well separated with respect to their variance.

\subsection{Order parameters}

We now introduce three order parameters that capture different aspects of transition configurations. The first, $q_1=n$, is the size $n$ of the largest cluster, which in the picture of classical nucleation theory is sufficient to describe nucleation processes. Fig.~\ref{fig:tse}(a) shows a scatter plot of cluster size $n(x)$ \emph{vs.} the committor probability $\pb(x)$ for all harvested configurations $x$. It shows the correct trend: for small $n$ the probability $\pb$ to commit to basin B is small while for large clusters $n$ it reaches unity. To assess the quality of $n$ as a reaction coordinate, we employ the log-likelihood maximization to extract $\al_0$ and $\al_1$ with $\al_2=\al_3=0$. From the condition $\pbm(n_\ast)=\tfrac{1}{2}$ we again obtain an estimate for $n_\ast$. We then collect all configurations with 
\begin{equation}
  \label{eq:tse}
  -0.25 \leqslant r \leqslant 0.25
\end{equation}
and plot the distribution of $\pb$ values for these configurations in Fig.~\ref{fig:tse}(f), which is rather different from the desired, peaked shape. Hence, while $n$ correctly describes the two basins, it is not sufficient to describe the TSE.

As a second order parameter $q_2=\rho$ we consider the density $\rho=n/a$ of the largest cluster. This order parameter describe how ``compact'' the cluster is and also reflects the density change between dilute and dense phase. While we already know the number of particles $n$, we also need to determine the area $a$ occupied by the cluster. To this end we employ the \emph{scipy.interpolate} library~\cite{scipy} using Gaussian basis functions to interpolate the density. The instantaneous interface is then formed by an appropriate density isocontour~\cite{will10}. We repeat the same analysis as for the cluster size shown in Figs.~\ref{fig:tse}(b) and~(g). Again, we find that the density is a good order parameter but not a good reaction coordinate.

\begin{figure}[b!]
  \centering
  \includegraphics{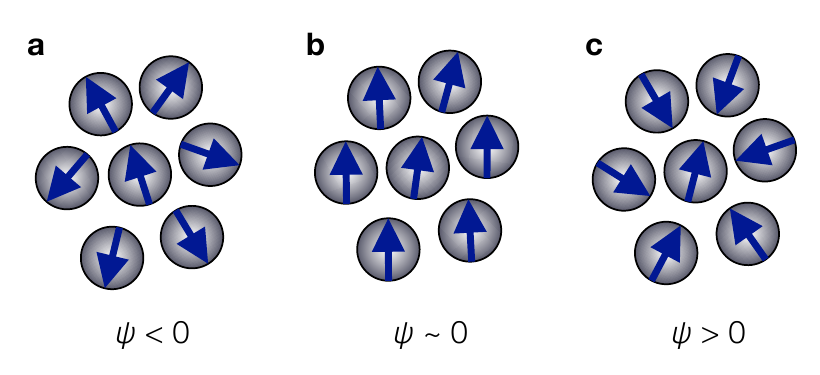}
  \caption{Illustration of the radial polarization Eq.~(\ref{eq:psi}). (a)~Orientations in a cluster point predominately outwards, leading to a negative value for $\psi$. (b)~If orientations mainly point in the same direction the cluster has a large net polarization but a small radial polarization. (c)~Only if orientations point inwards does the value for $\psi$ become positive.}
  \label{fig:polarization}
\end{figure}

As the final order parameter $q_3=\psi$, we employ the radial polarization
\begin{equation}
  \label{eq:psi}
  \psi(x) = -\frac{1}{n}\sum_{k\in C} \vec e_k\cdot(\x_k-\x_\text{cm}),
\end{equation}
where the sum runs over all $n$ indices $k\in C$ of particles that constitute the largest cluster $C$ and $\x_\text{cm}$ denotes the cluster's center of mass. This order parameter takes into account the orientations of the self-propelled disks. In contrast to the net polarization (the sum of orientations) it projects particle orientations onto the vector from the cluster center to the particle, with particles in the rim of the cluster acquiring a larger weight (due to their larger distance) than the inner particles. The behavior of $\psi$ is illustrated in Fig.~\ref{fig:polarization}. Again, we repeat the analysis as for the two previous order parameters, the result of which is shown in Figs.~\ref{fig:tse}(c) and~(h), with similar conclusions.

\subsection{Optimal reaction coordinate}

\begin{table}[b!]
  \centering
  \begin{tabular}{c|c|c|c|c|c}
    combination & $\ln L_n/\ln L$ & $\al_0$ & $\al_1$ & $\al_2$ & $\al_3$ \\
    \hline
    $n$, $\rho$, $\psi$ & 1.067 & -6.54 & 0.0099 & 3.30 & 0.82 \\
    $n$, $\psi$ & 1.065 & -4.50 & 0.0106 & & 0.97 \\
    $n$, $\rho$ & 1.038 & -10.41 & 0.0106 & 10.41 & \\
    $n$ & 1 & -3.41 & 0.0144 & & \\
    $\psi$ & 0.830 & -5.71 & & & 2.45 \\
    $\rho$ & 0.823 & -23.96 & & 30.85 &
  \end{tabular}
  \caption{Combination of order parameters sorted by their log-likelihood and normalized by the log-likelihood $\ln L_n$ of the cluster size. The combination of all three order parameters has the highest likelihood, but is only marginally better than the combination of cluster size $n$ with polarization $\psi$. Also given are the expansion coefficients $\al_i$.}
  \label{tab:lhood}
\end{table}

Table~\ref{tab:lhood} lists all possible linear combinations together with the normalized log-likelihood. Of the three single order parameters, the cluster size $n$ is the best approximation to a reaction coordinate. Since $r=0$ corresponds to the TSE, from the coefficients for the single order parameters we can calculate their transition value. For example, the critical cluster size is found to be $n_\ast=-\al_0/\al_1\simeq237$, which compares well with the direct average of $n_\ast\simeq232$.

Our analysis so far has revealed that all three order parameters, by themselves, are not good reaction coordinates. We now go a step further and test different combinations using the method of likelihood maximization, the result of which is included in Table~\ref{tab:lhood}. More complex models have a higher likelihood, however, the gain when using all three order parameters is marginal so that as the best model we consider the combination of cluster size $n$ and polarization $\psi$. For two order parameters, from $r=0$ we obtain a linear relation for the TSE, which is shown in Fig.~\ref{fig:tse}(d,e). We again apply the criterion Eq.~(\ref{eq:tse}) to identify the configurations constituting the TSE. The corresponding distributions $P(\pb|\text{TSE})$ shown in Figs.~\ref{fig:tse}(i) and~(j) are now indeed peaked around $\tfrac{1}{2}$ as desired. While the inclusion of the cluster density $\rho$ is already a significant improvement, the radial polarization $\psi$ is even more relevant.

Of course, the three order parameters taken into account here are not the only possible choices. Among others, we have also considered the anisotropy of clusters and the ratio of circumference. However, employing the described method of likelihood maximization, these three have been found to be the most relevant. This does not exclude the possibility that there is yet another order parameter that is superior in describing the nucleation of dense clusters of active Brownian disks.

\subsection{Role of polarization}

As mentioned, the cluster size $n$ is a natural choice that is often sufficient to describe nucleation. As a counterexample, for the crystallization of soft particles studied in Ref.~\citenum{lech11} it has been found that including the cluster surface improves the reaction coordinate. The physical picture for active Brownian particles is quite similar, also here the particles at the cluster's surface play a role. However, it is now their orientation that is important. This can be understood quite easily. If orientations at the surface point outwards [cf. Fig.~\ref{fig:polarization}(a)] they would leave the cluster, which then becomes instable and small clusters might even vanish. If the majority of particles at the surface points inwards [cf. Fig.~\ref{fig:polarization}(c)], these particles are blocked due to the inner particles, effectively stabilizing the cluster. Hence, the creation of a cluster is linked to a polarization of orientations, a collective fluctuation away from the expectation value of zero for active Brownian particles.

\begin{figure}[b!]
  \centering
  \includegraphics{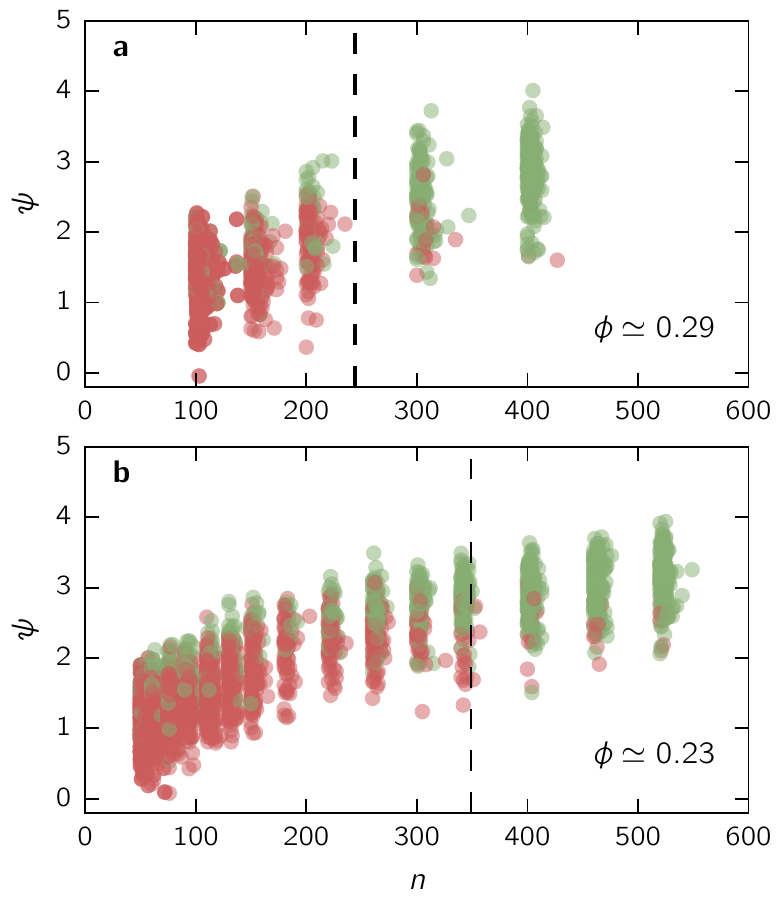}
  \caption{Scatter plot of cluster size $n$ and polarization $\psi$ for the FFS configurations for two packing fractions (a)~$\phi\simeq0.29$ and (b)~$\phi\simeq0.23$. The vertical lines indicate the critical cluster sizes $n_\ast$. The colors indicate whether the FFS trajectories started from this configuration have reached the next interface (green) or fallen back (red). Clearly, reactive trajectories originate in configurations with higher values of $\psi$.}
  \label{fig:ffs}
\end{figure}

To confirm this picture, we have analyzed in more detail configurations harvested with FFS. In Fig.~\ref{fig:ffs} we show all stored configurations with their cluster size $n$ and radial polarization $\psi$ for two densities. Configurations are colored according to whether they have reached the next interface, or have fallen back to the homogeneous suspension. Clearly, successful configurations typically have a larger value for $\psi$, indicating a higher polarization. Even beyond the critical cluster size $n_\ast$, large clusters fall back if their polarization is low.

\subsection{Droplet condensation/evaporation}

Computer simulations are necessarily performed in finite systems, which often has interesting and subtle consequences. For example, for liquid-vapor coexistence it has been found that in finite systems with linear dimension $L$ the homogeneous phase is stable also above the coexistence density $\phi_-<\phi<\phi_0$ and there is a condensation transition at a (slightly) higher packing fraction $\phi_0$, which in passive systems in two dimensions scales as $\phi_0-\phi_-\propto L^{-2/3}$~\cite{bind03}.

\begin{figure}[t]
  \centering
  \includegraphics{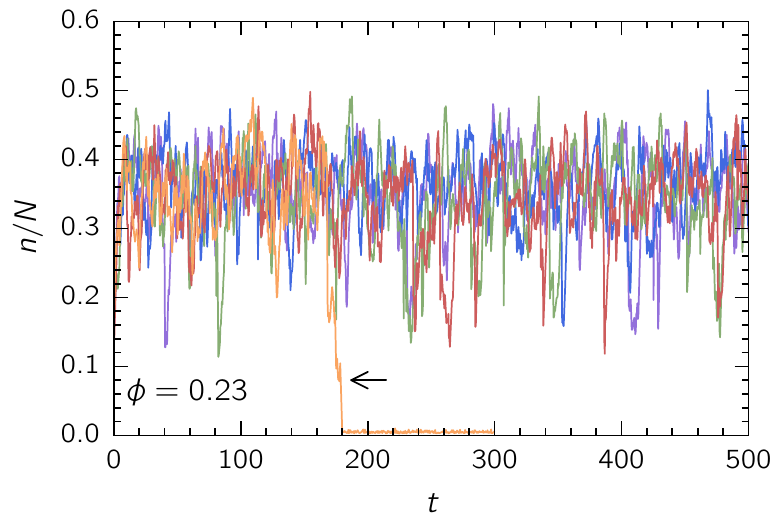}
  \caption{Cluster size $n$ as a function of time for the lowest packing fraction $\phi\simeq0.23$ studied. Of the 5 independent runs, in one run (arrow) the dense domain has melted and the suspension has become homogeneous again.}
  \label{fig:stability}
\end{figure}

It is straightforward to see that a similar picture should hold for active Brownian disks: the size of stable droplets is reduced when reducing the global packing fraction while the size of the critical nucleus increases. At some point small droplets cannot be stable anymore in finite systems. Interestingly, we find an ``echo'' of this evaporation transition showing that large fluctuations can destabilize clusters at low supersaturation with the suspension returning to the homogeneous state. To this end, we have run direct BD simulations at packing fraction $\phi\simeq0.23$ initialized with configurations harvested by FFS at the last interface, which have a numerical probability of unity to commit to the phase-separated state. Fig.~\ref{fig:stability} shows the time series of relative cluster sizes $n/N$ (the fraction of particles that make up the largest cluster). While clusters remain stable in most runs, we observe one run in which the cluster initially is stable but suddenly decays, after which the system remains homogeneous for a long time. This indicates that we are sufficiently close to the condensation/evaporation transition to observe fluctuations between homogeneous suspension and the droplet.


\section{Conclusions}

For active hard Brownian disks, we have studied the nucleation of dense domains using both Brownian dynamics and Forward Flux Sampling simulations. While active Brownian particles are driven away from thermal equilibrium due to self-propulsion, it has been found previously that their large-scale behavior can be described by an effective free energy~\cite{cate15,spec15}, linking their phase behavior to that of passive liquid-vapor separation. In particular, active hard Brownian disks undergo phase separation for sufficiently large propulsion speeds. Here we have shown that also the phase-separation kinetics is qualitatively very similar: the nucleation rate is exponentially small close to the binodal and increases with increasing density. At the same time the critical size of domains (clusters) leading to phase separation decreases. This implies an effective description in terms of a barrier that needs to be overcome by rare and collective fluctuations. Performing a committor analysis we have found that the orientations of particles in the rim of clusters play an important role. In agreement with the idea of a ``spinodal'' marking the loss of linear stability, going to high supersaturation the barrier becomes low and there is a qualitative change from nucleation to demixing, \emph{i.e.}, domains appear almost instantaneously and coarsen over time. For the density dependence $n_\ast(\phi)$ of critical cluster sizes we found an exponent $3/2$ in contrast to $2$ for passive systems. Still, the kinetics of phase-separating active Brownian disks is surprisingly well described by equilibrium concepts. We even find evidence that close to coexistence droplets evaporate in finite systems.

The nucleation scenario unravelled in our simulations can be confirmed by experiments on light-activated diffusiophoretic colloids which exhibit phase separation~\cite{butt13}. In principle, experimental information on the cluster kinetics is directly available and the activity can be conveniently tuned to initiating the nucleation process. Future work should address the impact on hydrodynamic interactions~\cite{zott14,mata14} on phase separation kinetics.  Moreover, our findings provide a starting point to construct a microscopic theory for nucleation, where the effective free energy functional is employed and the relevant order parameters identified here play a crucial role~\cite{luts12}.


\acknowledgements

We gratefully acknowledge Julian Bialk\'e for preliminary results, Jonathan T. Siebert for stimulating discussions, and ZDV Mainz for computing time on the MOGON supercomputer. We acknowledge financial support by the DFG through collaborative research center TRR 146 (D.R.) and within priority program SPP 1726 (H.L. and T.S., grant numbers LO 418/17-1 and SP 1382/3-1).


\end{document}